\documentclass[iop,apj]{emulateapj}
\usepackage{amsmath,amssymb,amstext}
\pdfoutput=1

\usepackage[breaklinks,colorlinks,citecolor=blue,linkcolor=magenta]{hyperref}

\usepackage[all]{hypcap}

\hypersetup{pdfauthor={Yao-Yuan Mao et al.},pdftitle={The Dependence of Subhalo Abundance on Halo Concentration}}
\shorttitle{Subhalo Abundance and Halo Concentration}
\shortauthors{Mao, Williamson, \& Wechsler}

\newcommand*{\myplotpath}{plots}

\newcommand*{\sub}[2]{{#1}_{\text{#2}}}
\newcommand*{\vmax}{\sub{v}{max}}
\newcommand*{\vpeak}{\sub{v}{peak}}
\newcommand*{\Vmax}{\sub{V}{max}}
\newcommand*{\Vvir}{\sub{V}{vir}}
\newcommand*{\vcut}{\sub{V}{cut}}
\newcommand*{\mvir}{\sub{M}{vir}}
\newcommand*{\nsub}{\sub{N}{sub}}
\newcommand*{\nsat}{\sub{N}{sat}}
\newcommand*{\kms}{km$\,$s$^{-1}$}

\DeclareMathOperator{\Pois}{Pois}

\newcommand{\add}[1]{#1}

\begin{document}

\title{The Dependence of Subhalo Abundance on Halo Concentration}
\author{Yao-Yuan Mao\altaffilmark{1}, Marc Williamson, and Risa H.~Wechsler\altaffilmark{2}}
\affil{Kavli Institute for Particle Astrophysics and Cosmology
  \& Physics Department, Stanford University, Stanford, CA 94305, USA}
\affil{SLAC National Accelerator Laboratory, Menlo Park, CA, 94025, USA}
\altaffiltext{1}{\href{mailto:yymao@stanford.edu}{yymao@stanford.edu}}
\altaffiltext{2}{\href{mailto:rwechsler@stanford.edu}{rwechsler@stanford.edu}}

\begin{abstract}
Hierarchical structure formation implies that the number of subhalos
within a dark matter halo depends not only on halo mass, but also on
the formation history of the halo.  This dependence on the formation
history, which is highly correlated with halo concentration, can
account for the super-Poissonian scatter in subhalo occupation at a
fixed halo mass that has been previously measured in simulations.
Here we propose a model to predict the subhalo abundance function for
individual host halos, that incorporates both halo mass and
concentration.  We combine results of cosmological simulations with a
new suite of zoom-in simulations of Milky Way-mass halos to calibrate
our model.  We show the model can successfully reproduce the mean and
the scatter of subhalo occupation in these simulations.  The
implications of this correlation between subhalo abundance and halo
concentration are further investigated.  We also discuss cases in
which inferences about halo properties can be affected if this
correlation between subhalo abundance and halo concentration is
ignored; in these cases our model would give a more accurate
inference.  We propose that with future deep surveys, satellite
occupation in the low-mass regime can be used to verify the existence
of halo assembly bias.
\end{abstract}

\keywords{dark matter ---  galaxies: halos --- methods: analytical --- methods: numerical}

\maketitle

\section{Introduction}

Bridging our understanding of the processes of galaxy formation and of
the evolution of dark matter halos remains one of the primary
challenges in modern cosmology. While $N$-body simulations provide
detail about the formation and evolution of dark matter halos, it is
still observationally challenging to directly probe their properties.
Nevertheless, extensive work over the past decade has used
observations of galaxy's spatial distributions to constrain models of
the galaxy--halo connection, which reveals how galaxies form in
halos~\citep[e.g.,][]{2002ApJ...575..587B,2011ApJ...736...59Z,2013ApJ...771...30R}.
As new observations become more precise, it is crucial to understand
possible systematic uncertainty and bias in those models.

The two main characteristics of a dark matter halo are its mass,
usually calculated by setting a spherical over-density region, and its
formation history. The latter is also highly correlated with the
density profile of the halo, and hence with the concentration and with
the maximal circular velocity $\vmax$ of the
halo~\citep{2002ApJ...568...52W}. Halos of the same mass but different
formation history can have very different characteristics or reside in
different
environments~\citep[e.g.,][]{2001MNRAS.321..559B,2006MNRAS.367.1781A,2007MNRAS.378...55M}.

The abundance of subhalos within a dark matter halo most strongly
correlates with the mass of the halo
\citep[e.g.,][]{2004ApJ...609...35K}.  Nevertheless, at a fixed halo
mass, the subhalo abundance also correlates with the formation history
of the halo~\citep{2005ApJ...624..505Z,2006ApJ...639L...5Z,2009ApJ...696.2115I}. This
correlation, despite its significance in modeling satellite
occupation, is often neglected, mostly because it does not manifest
itself when the Poisson scatter is comparable to the number of subhalos in consideration.
Satellite occupation, or richness, is often used as a proxy of host
halo mass, especially for optical observations of
clusters~\citep{2009ApJ...703..601R,2010ApJ...708..645R}.  
The scatter in the mass distribution inferred from richness can be 
underestimated if this correlation with concentration is neglected.

In this work, we investigate again the correlation between subhalo
abundance and halo concentration, and propose a simple model that
describes this correlation. This model can also be used to extend the
subhalo abundance function for a given host halo beyond the resolution
limit, and enables us to evaluate how this correlation may manifest in
a range of observable statistics.

The simplest approach to extend the subhalo abundance function beyond
the resolution limit is to extrapolate a parametrized subhalo
abundance function.  The subhalo abundance function is most commonly
modeled by a power law, and the parameters of the model can be
calibrated against simulations.  Studies have shown this method
describes the subhalo abundance functions in $N$-body simulations very
well~\citep{2004MNRAS.355..819G,2004ApJ...609...35K,2008MNRAS.387..689G,2008MNRAS.391.1685S,2009MNRAS.399..983A,2010MNRAS.406..896B,2013ApJ...767..146I,2014MNRAS.445.1820C},
at least for host halos in a narrow mass range.

In order to calibrate this kind of model over a wide range
of mass, usually a suite of cosmological simulations and zoom-in
simulations is needed. Zoom-in simulations, such as the Aquarius and
Phoenix simulations~\citep{2008MNRAS.391.1685S,2012MNRAS.425.2169G},
are particularly powerful for measuring subhalo abundance function
at high resolution but still with reasonable costs. However, if
one wants to study the halo-to-halo scatter in the subhalo abundance
function, a fairly large sample size is required. More recently,
two re-simulation suites have been completed with tens to hundreds of
simulations in specific small mass ranges: the Rhapsody
\citep[cluster-mass halos,][]{2013ApJ...767...23W} and ELVIS simulations
\citep[Milky Way-mass halos,][]{2014MNRAS.438.2578G}.

While these fitting models can usually describe simulations
fairly well, they often capture the minimal relevant physics
for the particular questions that are being addressed. 
A more elaborate approach is to consider the assembly histories of dark matter halos and
the evolution of halo mass function~\citep{2011ApJ...741...13Y}.
One can further consider more relevant subhalo dynamics
when modeling subhalo abundance beyond the resolution limit
by  tracking the orbits of subhalos and adding subhalos that do not appear or are disrupted in
simulations~\citep{2005ApJ...624..505Z,2014arXiv1403.6827J,2014arXiv1403.6835V}.
Instead of fitting the abundance function, 
this kind of approach considers most physical details,
but at the same time can be more difficult to constrain.

In this work, we focus on an empirical model which directly uses mass
and $\vmax$ of the host halo to predict the subhalo abundance function,
and calibrate the model against cosmological and zoom-in simulations.
This model is essentially the simplest possible model of subhalo abundance
function that takes halo formation history into account. In principle,
a more sophisticated model (i.e., models that track subhalo evolution)
could produce similar results. However, our simple model provides
a straightforward way to evaluate this correlation between subhalo
abundance and halo formation history, and to evaluate its implications
for various observables.

This paper is organized as follows.  In \autoref{sec:sims} we describe
the simulations used in this study.  In \autoref{sec:model} we first
discuss the correlation between subhalo abundance and halo formation
history, and then we describe and calibrate the model which predicts
the subhalo abundance.  In \autoref{sec:discussion}, we further
discuss the implications of this correlation between subhalo abundance
and halo concentration.  We summarize this paper in
\autoref{sec:summary}.

\section{Simulations} \label{sec:sims}

In this study we use a cosmological simulation \texttt{c125-2048} and
also present a new set of zoom-in simulations of Milky Way-mass halos.

The \texttt{c125-2048} box\footnote{\label{fn:c125}Provided by Matthew Becker (M.~Becker et al.~2015, in preparation)} is a dark matter-only cosmological simulation run
with \textsc{L-Gadget}~\citep[based on \textsc{Gadget-2},][]{2001NewA....6...79S,2005MNRAS.364.1105S}. 
The box has $2048^3$ particles and a side
length of 125 Mpc$\,h^{-1}$, resulting in a particle mass of $1.8
\times 10^7 M_\odot h^{-1}$. The softening length used is
0.5 kpc$\,h^{-1}$, constant in comoving length. The cosmological
parameters are $\Omega_m=0.286$, $\Omega_\Lambda=0.714$, $h=0.7$, $\sigma_8=0.82$, and $n_s=0.96$.
The initial conditions are generated by
\textsc{2LPTic}\footnote{\url{http://cosmo.nyu.edu/roman/2LPT/}}~\citep{2006MNRAS.373..369C}
at $z=199$, with the power spectrum generated by
\textsc{Camb}.\footnote{\url{http://camb.info/}}

\begin{figure*}
  \centering \includegraphics[width=\textwidth]{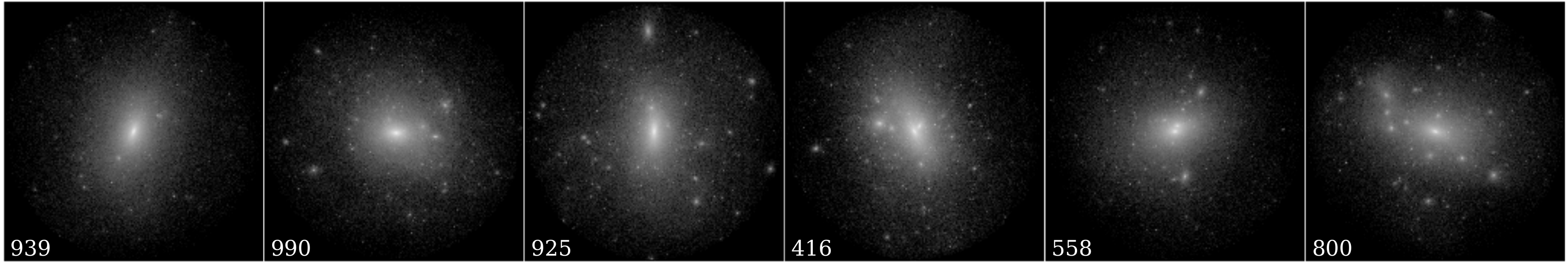}
  \caption{Images of the zoom-in simulations of six Milky Way-mass halos, 
  from our suite of 46 halos. The concentration of these selected halos decreases 
  from left to right.}
  \label{fig:mw_halos}
\end{figure*}

\begin{figure}
  \centering \includegraphics[width=\columnwidth]{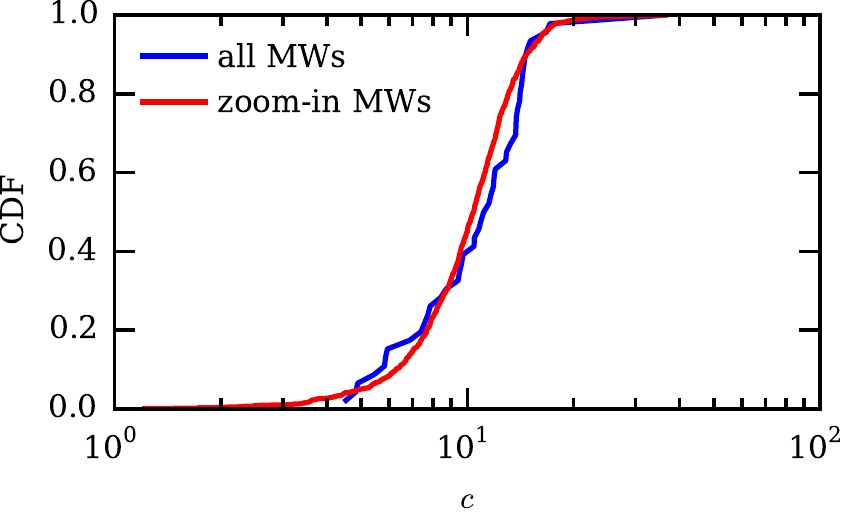}
  \caption{The cumulative distribution of concentration (in log scale)
    for the zoom-in Milky Way halos (red) and all the halos in the same mass
    range in the\texttt{c125-1024} box (blue).}
  \label{fig:mw_c_dist}
\end{figure}

The new suite of zoom-in simulations consists of 46
Milky Way-mass halos, selected from the \texttt{c125-1024} box (see footnote~\ref{fn:c125}), 
which is a low-resolution version of the \texttt{c125-2048} box.
The parameters and initial conditions of these two boxes are identical, but
\texttt{c125-1024} contains only $1024^3$ particles and starts at $z=99$. All the selected
halos fall in the mass range $\mvir=10^{12.1\pm 0.03}M_\odot$ in the
\texttt{c125-1024} box. The initial conditions of these zoom-in
simulations are generated with the publicly available \textsc{Music}
code\footnote{\url{https://bitbucket.org/ohahn/music}}~\citep{2011MNRAS.415.2101H},
and are matched to the cosmological box up to the $1024^3$
scale.
The Lagrangian volume where the highest-resolution particles are placed is
set by the rectangular volume which the particles within $10 \sub{R}{vir}$ 
of the present-day halo occupied at $z=99$.
The mass of the highest-resolution particles in the zoom-in
simulations is $3.0 \times 10^5M_\odot h^{-1}$. The softening length
in the highest-resolution region is 170 pc$\,h^{-1}$ comoving.
\add{\autoref{fig:mw_halos} shows the images of 6 of these zoom-in simulations.}
\autoref{fig:mw_c_dist} compares the concentration distribution of
this sample of Milky Way-like halos with the full sample in the mass range in
the \texttt{c125-2048} box. The concentration distribution of the
selected sample is slightly wider than that of all the host halos in
the mass range.

In the analysis, we use \textsc{Rockstar}\footnote{\url{https://bitbucket.org/gfcstanford/rockstar}}
for halo finding and \textsc{Consistent Trees}\footnote{\url{https://bitbucket.org/pbehroozi/consistent-trees}}
for tree building~\citep{2013ApJ...762..109B,2013ApJ...763...18B}.
The halos are defined with $\sub{\Delta}{vir} \simeq 99.2$ for this cosmology.
Subhalos are defined as halos that are within $\sub{R}{vir}$ of any
other larger halo. Halos that are not a subhalo are called
\textit{host halos} throughout this paper.

The particle mass of a simulation cannot be directly translated into the
maximal circular velocity, $\vmax$, to which the simulation converges. By
inspecting the velocity function, we estimate that a conservative lower
limit for the convergence of the \texttt{c125-2048} box is 40 \kms{},
and that of the zoom-in Milky Way simulations is 9 \kms{}.

\section{Modeling Subhalo Abundance}
  \label{sec:model}

In this section, we present a framework to model the subhalo abundance
of individual host halos. We first discuss the correlation between
subhalo abundance and host halo concentration, and observe
qualitatively how host halo concentration affects subhalo abundance
function.  We further argue that for a given host halo, the number of
subhalos is consistent with a Poisson distribution.  Then we describe
both the framework and the specific parameterization of our model, and
calibrate the model against the aforementioned simulations.  Finally
we briefly discuss the universality of the subhalo abundance function.

\subsection{Dependence of Subhalo Abundance on Halo Concentration}

$N$-body simulations have shown that the subhalo abundance function
averaged over a sample of host halos of a similar mass approximately
follows a power law, and its form is nearly universal for different
host halo masses when scaled
properly~\citep[e.g.,][]{2004MNRAS.355..819G,2004ApJ...609...35K,2010MNRAS.406..896B}.
Hence, the simplest model of subhalo abundance is to describe the mean
number of subhalos, $\langle \nsub \rangle$, as a function of host
halo mass only.  Although this simple kind of model can predict the
mean number of subhalos at a given host halo mass in simulations
fairly well, it cannot explain the dependence of subhalo abundance on
host halo concentration, as shown in \citet{2005ApJ...624..505Z}.

\begin{figure}
  \centering \includegraphics[width=\columnwidth]{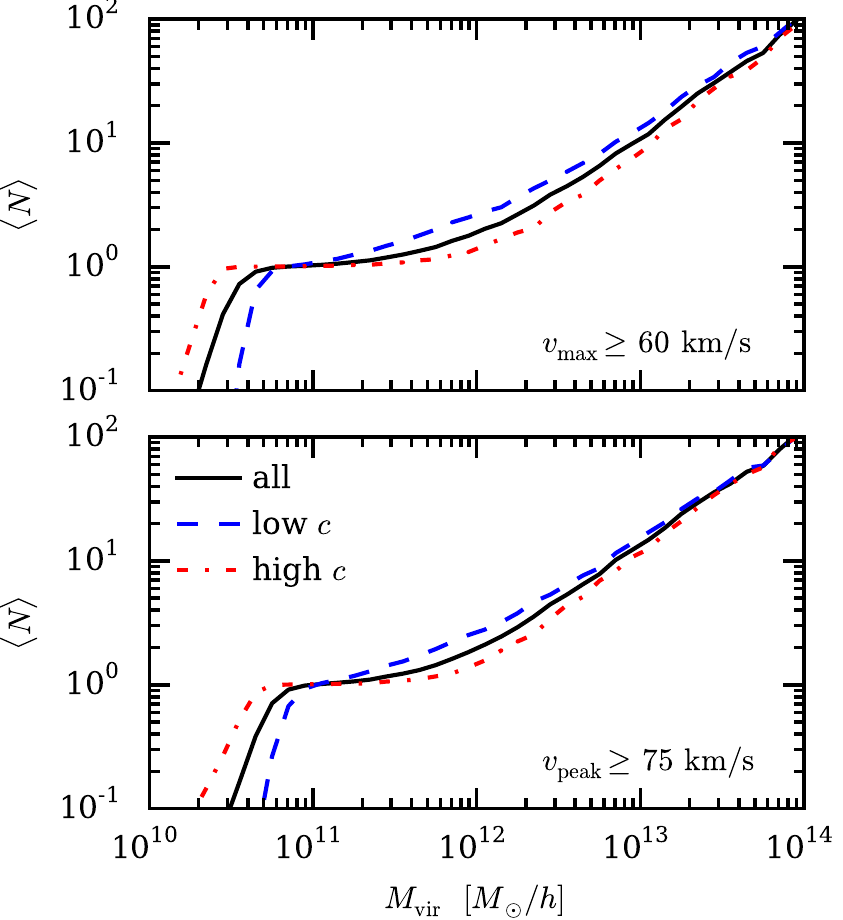}
  \caption{Number of \emph{galaxies}, i.e., halos (including both hosts and subhalos)
    with a cut in $\vmax$
    (upper) or in $\vpeak$ (lower), as a function of host halo mass.
    The black solid line shows all host halos, while the blue dashed line
    and the red dash--dot line show the host halos with the lowest and the
     highest 25\% of concentration, respectively.}
  \label{fig:hod}
\end{figure}

To see how host halo concentration affects the number of subhalos, in
\autoref{fig:hod} we plot the mean number of halos (including hosts
and subhalos) whose $\vmax$ (or $\vpeak$) is larger than 60 \kms{} (or
75 \kms{}) as a function of host halo mass. We plot this relation for
all the host halos and for only halos with the highest and the lowest
25\% of concentration in each mass bin. We can clearly see that halos
of high concentration tend to have fewer subhalos, and also see that
this is not a small effect, especially when the halo halo mass is
about $10^{12} M_\odot h^{-1}$. We note that at higher host halo mass,
this difference becomes smaller because high-mass halos have a smaller
spread in concentrations than low-mass halos.

\begin{figure}
  \centering \includegraphics[width=\columnwidth]{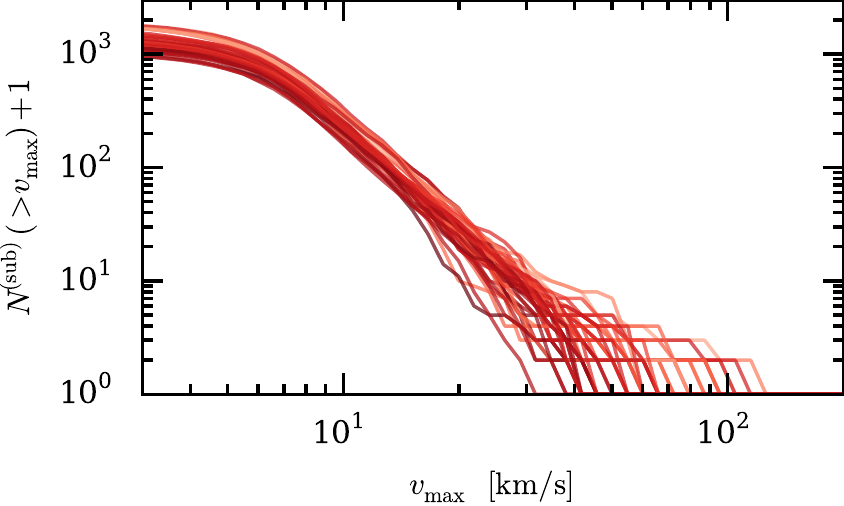}
  \caption{The subhalo $\vmax$ function for the 46 zoom-in simulations
    of the Milky Way halos. Each line represents one host halo and is colored
    according to the ratio $\Vmax/\Vvir$ of the host halo.
    \add{Darker color reprensents halos of higher concentration (larger $\Vmax/\Vvir$). The gray band on the left shows the regime affected by
      resolution, where the abundance function bends due to unresolved subhalos.}}
  \label{fig:slope_MW}
\end{figure}

We now take a closer look at how concentration affects the subhalo
abundance on a halo-by-halo basis for host halos of the same mass. In
\autoref{fig:slope_MW}, we plot the subhalo $\vmax$ function for all the
zoom-in simulated Milky Way-mass halos.
The subhalo $\vmax$ functions in \autoref{fig:slope_MW} are colored
according to the concentration of their respective host halos.
We observe two prominent features:
\begin{enumerate}
   \item All these halos fall in a very narrow mass bin (smaller than
       0.08 dex), yet there is a significant halo-to-halo scatter in
       their subhalo $\vmax$ functions. The halo-to-halo scatter seems
       to affect mostly the normalization of the subhalo $\vmax$
       function, and the trend roughly follows the concentration
       trend, which is indicated in colors --- darker lines sit lower.
  \item On the log--log plot, subhalo $\vmax$ functions are mostly
      parallel to one another, especially in the regime where $\nsub >
      10$.  This suggests the power-law index is roughly a constant
      from halo to halo\add{. Also, for each individual halo, 
      the deviation of the abundance function from a simple power law}
      is much smaller than the halo-to-halo
      scatter when $\nsub$ is large.
\end{enumerate}
In \citet{2013ApJ...767...23W}, the authors also find that
the numbers of subhalos in different $\vmax$ bins are correlated, 
especially when $\nsub$ is large.
This agrees with our findings here.

This correlation between the subhalo number and host halo
concentration has been found and discussed in, for example,
\citet{2005ApJ...624..505Z}, \citet{2011ApJ...738...22W}.  This correlation can be understood by
the hierarchical formation of halos: conditioned on a fixed halo mass,
halos with higher concentration form early, and subhalos in these
halos are stripped longer to a lower mass and $\vmax$, and
some could already be completely disrupted and merged with the host.
Both effects would result in a smaller number of subhalos at a fixed velocity cut.

\subsection{\add{Small-scale} Poisson Scatter}

It is also known and shown explicitly by \citet{2010MNRAS.406..896B}
that the scatter in the number of subhalos is super-Poissonian when
the mean number is much larger than 1, The authors argue this
super-Poissonian scatter is a sum of a Poisson scatter and an
\emph{intrinsic} scatter \citep[see also related discussion in][]{2011ApJ...743..117B}.  

Here we further claim that the Poisson scatter should exist on a
single-halo basis. 
\add{That is, given a host halo and its environment, 
the small-scale variation would result in a Poisson scatter in its 
subhalo abundance.
On the other hand, the \emph{intrinsic} scatter (or more
precisely called the \emph{halo-to-halo} scatter) is then in principle
all possible scatter among host halos}.

To verify that the subhalo abundance function is always subject to
\add{this small-scale} Poisson scatter 
\add{when} we consider a single host halo, i.e.,
\begin{equation}
  (\nsub\vert\,\text{host}) \sim \Pois(\langle \nsub\vert\,\text{host} \rangle),
\end{equation}
we run 13 zoom-in simulations of a single halo,
with different random seeds for the small-scale modes. 
All these 13 realizations have the same simulation setup as described above, 
and also the same large-scale initial conditions down to the scale
of $k \sim 16.4\,h\,\text{Mpc}^{-1}$, 
which is equivalent to $2048^3$ particles in the box.
This scale roughly corresponds to a host halo mass of $2.5 \times 10^{10} M_\odot h^{-1}$, 
or host $\Vmax \sim 50$ \kms{}.

\begin{figure}
  \centering \includegraphics[width=\columnwidth]{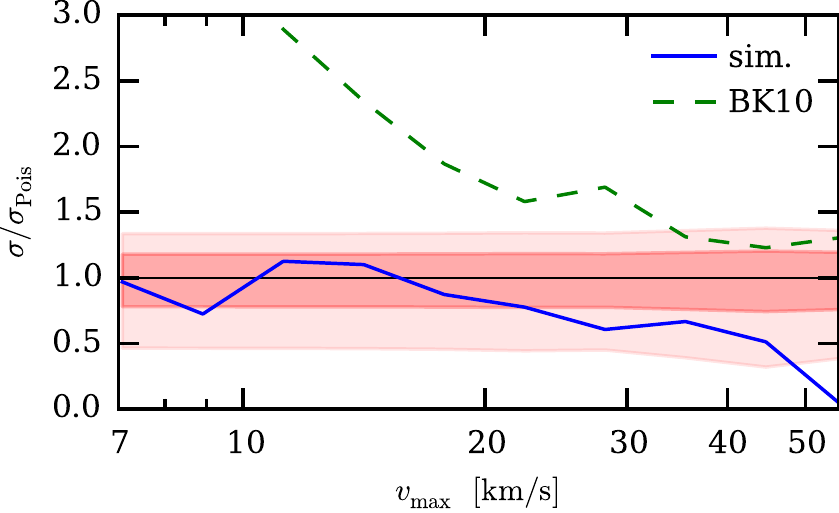}
  \caption{The blue line shows
    $\sigma/\sub{\sigma}{Pois}$ in bins of $\vmax$, calculated over
    the 13 halos of the same large-scale initial conditions.
    The red bands show the 1-$\sigma$ (dark) and 2-$\sigma$ (light) confidence
    interval if $N$ follows a Poisson distribution and given that there are
    13 samples.
    The green dashed line shows the super-Poissonion scatter~\citep[Figure~8]{2010MNRAS.406..896B}
    for comparison.}
  \label{fig:poisson}
\end{figure}

\autoref{fig:poisson} shows $\sigma/\sub{\sigma}{Pois}$, where
$\sigma$ is standard deviation and $\sub{\sigma}{Pois}=\sqrt{\langle N
\rangle}$, i.e., the square-rooted ratio of the variance to the mean of
the number of subhalos, in bins of $\vmax$ of the subhalos. The
variance and the mean are calculated over the 13 halos of the same
large-scale initial conditions. If the number of subhalos in a given
$\vmax$ bin follows a Poisson distribution, this ratio would be 1. In
\autoref{fig:poisson}, one can see that at higher values of $\vmax$,
this ratio is less than 1, which is expected due to the constrained
large-scale modes.  At smaller $\vmax$, this ratio approaches 1.
Although the sample size is small, the typical number of subhalos
above $\vmax=10$ \kms{} is already more than 200. 
Hence, if the super-Poissonian scatter truly exists
\add{at the scales within a single host halo},
one would expect the ratio to be larger than 1 at small $\vmax$, 
scaling similar to the green dashed line, which includes the super-Poissonian scatter. 
This test suggests that, for a \emph{given} host halo (and its environment), the
scatter of its subhalo abundance is consistent with Poisson scatter.
The super-Poissonian scatter in a fixed host halo mass
cannot solely come from small-scale modes, and should
be a result of the scatter in the host halo properties at that
fixed mass, combined with dependence of the subhalo abundance on these
properties.

\subsection{Framework of the Model}

Now we present the framework of our subhalo abundance model. We first
outline our model that describes the number of subhalo for a given host
halo, and the parameters of the model. Then we further present how to
relate these model parameters to the properties of the host halos. In this
fashion, we can clearly separate the Poisson scatter in each individual
host halo from the halo-to-halo scatter.

Mathematically, we can model the subhalo abundance function as a counting
process. Here the counting process we consider is counting over the proxy
variable (i.e., $\vmax$ or $\mvir$), not over the physical time. Although
the mathematical term we used is \textit{process}, we are not considering
the physical evolution of the subhalo merging process, but only the
number of subhalos at a given time.

Let $N(v)$ denote the number of subhalos whose $\vmax$ (or other proxy,
which for simplicity we call $v$) is greater than or equal to $v$. Note
that $N(v)$ is always an integer and has the following properties:
\begin{subequations}
\begin{align}
  &N(v_1) \geq N(v_2) \text{, if } v_1 \leq v_2, \\
  &N(v) = 0 \text{, if } v \geq \vcut,
\end{align}
\end{subequations}
where $\vcut$ is a scale above
which there are no subhalos. The value of $\vcut$ depends on the host halo.

We further argue that this counting process is an inhomogeneous Poisson
process. That is, the number of subhalos in the interval $[v_1, v_2)$
follows a Poisson distribution and is independent of the counts in any
other disjoint intervals. We can write
\begin{equation}
  \left[N(v_1) - N(v_2) \right] \sim \Pois(\lambda(v_1, v_2)),
\end{equation}
and
\begin{equation}
  \label{eq:rate}
  \lambda(v_1, v_2) = \left(\frac{v_1}{V_0}\right)^n
    - \left(\frac{\min (v_2, \vcut)}{V_0}\right)^n,
\end{equation}
where $V_0$ is a positive parameter and $n$ is a negative parameter,
and both could depend on the host halo.
Note that the parameters $V_0$ and $\vcut$ should have the unit of the proxy.
For example is the proxy is $\vmax$, they should have the unit of velocity.
If one uses $\mvir$ instead as the proxy, they should have the unit of mass.

The expected number of subhalos whose $\vmax \geq v$ is then simply
\begin{equation}
  \label{eq:mean}
  \langle N(v) \rangle = \left(\frac{v}{V_0}\right)^n
    - \left(\frac{\vcut}{V_0}\right)^n.
\end{equation}

We note that by introducing the $\vcut$ scale, we do not need an
additional exponential cutoff in the model. The average subhalo
abundance function naturally drops off at the high end, and resembles
a exponential cutoff. There are two strengths of this approach. First,
the parameter $\vcut$ has a clear physical meaning; no subhalo can
have $\vmax$ (or any proxy in use) that is larger than $\vcut$.
Second, when implementing this model, one does not need to worry about
the chance of having a subhalo with a very large $\vmax$. The chance
of having such an outlier is remote but still finite when using an
exponential cutoff, while in our model the probability of a subhalo
with $\vmax \geq \vcut$ is zero by construction.

With our framework, there is a straightforward algorithm to create a
set of values which represents the set of the subhalo $\vmax$ values 
of a particular host halo, given a known threshold $\sub{v}{thres}$. 
This algorithm helps to generate a mock catalog of subhalos
beyond the resolution limit.
To generate this set, one first draws one random number
$k$ from a Poisson distribution of mean $N(\sub{v}{thres})$ according to
Equation~\eqref{eq:mean}, with $\sub{v}{thres}$ being the minimal possible $\vmax$ value
in the desired set.  Then one draws $k$ random numbers $X_1,
\dots, X_k$ from a uniform distribution $\mathcal{U}(0,1)$. 
The desired set would then be $\{f(X_1), \dots, f(X_k)\}$, where
\begin{equation}
f(x) := V_0 \left[N(\sub{v}{thres}) \cdot x + \left(\frac{\vcut}{V_0}\right)^n\right]^{1/n}
\end{equation}
is the inverse function of Equation~\eqref{eq:mean}.

\subsection{Calibrating the Model}
  \label{sec:calibrating}
  
So far we have introduced three parameters that are associated with
the host halo: $\vcut$, the largest scale a subhalo could have; $V_0$,
the overall normalization of the subhalo abundance function; and $n$,
the power-law index (log--log slope) of the subhalo abundance function.
In principle, the values of these three parameters in different host
halos do not need to follow any universal relation, and can depend on
\emph{any} host halo property. Nevertheless, since the dark matter
halos in dissipationless simulations do have many universal
properties, it is plausible that some universal relations relating
these three parameters to the host halo properties would already make
a good approximation.

For conventional models that describe $\langle N \rangle$ as a
function of host halo mass only, one can parameterize the variables in
Equation~\eqref{eq:rate} as follows
\begin{subequations}
  \label{eq:para_const}
\begin{align}
    V_0 &= a \, \Vvir , \\
    \vcut &= b \, \Vvir, \\
    n &= n_0,
\end{align}
\end{subequations}
where $\Vvir$ refer to the circular velocity at $\sub{R}{vir}$ of the host halo,
$a$, $b$, and $n_0$ are all constants that do \emph{not} depend on
any host halo properties.

However, we already know that the parameterization above cannot account
for the dependence on halo concentration.
Here we present a specific parameterization that replaces $a$ and $b$
in Equations~\eqref{eq:para_const} with functions of $(\Vmax/\Vvir)$.
Particularly, we set
\begin{subequations}
  \label{eq:para_func}
\begin{align}
    a &:= a_0 \, \left(\frac{\Vmax}{\Vvir}\right)^\alpha , \\
    b &:= b_0 \, \left(\frac{\Vmax}{\Vvir}\right)^\beta,
\end{align}
\end{subequations}
where $a_0$, $b_0$, $\alpha$, and $\beta$ are constant.
Here $\Vvir$ and $\Vmax$ refer to the host halo, and their ratio
can be viewed as a proxy of the halo concentration or formation time.
When $\alpha=\beta=0$, this falls back to the conventional model which has
no concentration dependence.

With this particular parametrization which incorporates host halo
concentration, we can calibrate the model against simulations. With
the \texttt{c125-2048} box, we find the values listed in
\autoref{table:para} provide decent \add{descriptions} to both the mean and the
scatter of subhalo abundance across a wide range of mass.  We also
find the values for two different redshifts ($z=1$ and 3) and for
using $\vpeak$ as the proxy.  Note that if one use $\vpeak$ as the
proxy instead of $\vmax$, the dependence on concentration is slightly
weaker (see the values of $\alpha$ in \autoref{table:para}).

\capstartfalse
\begin{deluxetable}{ccccccc}
\tablecaption{Parameter Values\label{table:para}}
\tablehead{
  \colhead{Proxy} & \colhead{Redshift} & \colhead{$a_0$} & \colhead{$\alpha$} & \colhead{$b_0$} & \colhead{$\beta$} & \colhead{$n_0$}
}
\startdata
$\vmax$  & 0 & 0.49 & $-0.9$ & 1.4 & $-2.5$ & $-2.90$ \\
$\vmax$  & 1 & 0.85 & $-1.0$ & 1.4 & $-1.0$ & $-2.80$ \\
$\vmax$  & 3 & 1.70 & $-1.0$ & 1.4 & $-0.8$ & $-2.60$ \\
$\vpeak$ & 0 & 0.67 & $-0.8$ & 1.4 & $-2.5$ & $-2.75$
\enddata
\tablecomments{See Equations~\eqref{eq:para_const} and \eqref{eq:para_func}
  for the definitions of these parameters. \add{See text of Section~\ref{sec:calibrating} for details.}}
\end{deluxetable}
\capstarttrue

\begin{figure*}
  \centering \includegraphics[width=\textwidth]{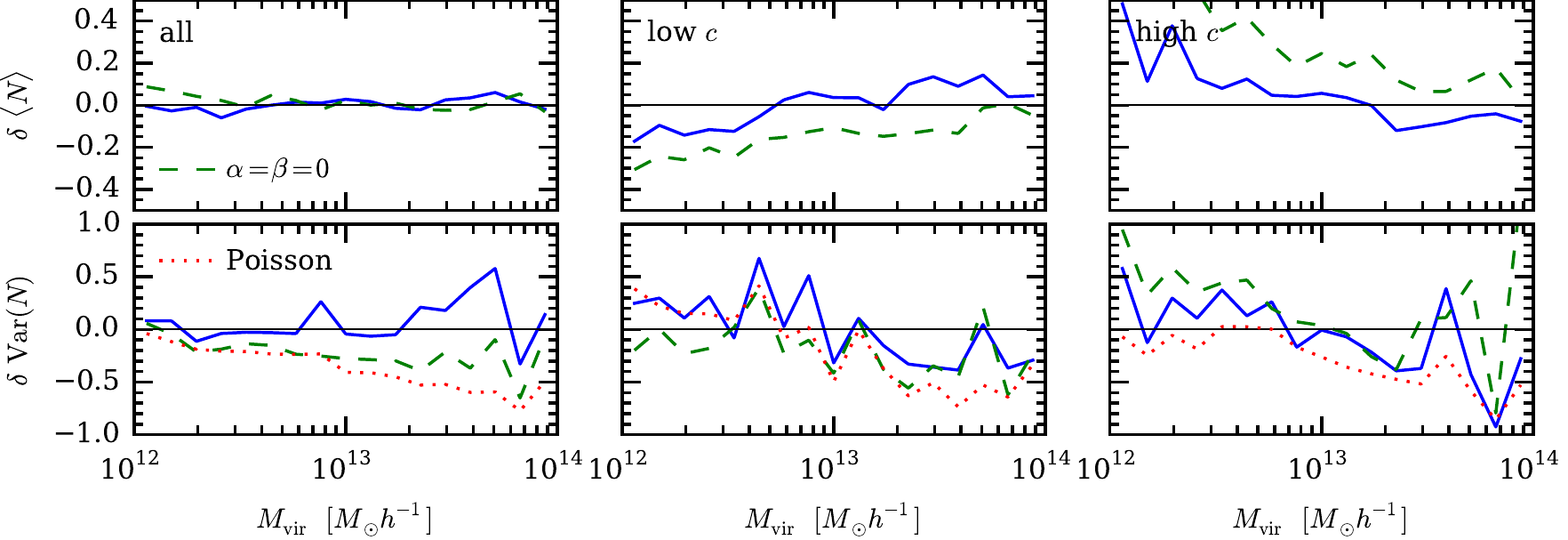}
  \caption{Relative difference
    between the model prediction and simulation of mean (upper row)
    and of variance (lower row) of the number of subhalos, in bins of
    host halo mass. The middle and the right columns show the lowest
    and the highest 25\% of concentration, respectively. Blue solid
    line shows the model we present here. The green dashed line is a
    model that depends only on host halo mass (i.e., $\alpha=\beta=0$).
    The red dotted line shows the Poisson scatter given the
    mean value in each bin.}
  \label{fig:model}
\end{figure*}

\autoref{fig:model} compares \add{simulations with} the prediction from this model \add{with the parameters listed in \autoref{table:para}}.
\add{In the simulations, we bin host halos according to their mass, 
in a wide range of masses ($10^{12}$--$10^{14} M_\odot h^{-1}$),
and measure the mean and variance of number of subhalos whose 
$\vmax > 50$ \kms{} in each bin.
For each host halo we also predict the number of subhalos with the model,
and measure the binned mean and variance in the same way as with simulations.
Then we plot the relative difference between the model prediction and the simulation as a function of host halo mass in \autoref{fig:model}. 
The relative difference is defined as 
$\delta X := X_\text{model}/X_\text{sim} - 1$, where $X$ could be the
mean (upper panels) or variance (lower panels) of number of subhalos
in each mass bin.}

\add{As \autoref{fig:model} shows,} our model can reproduce the
mean and variance of the number of subhalos in all mass bins very
well.
\add{We also plot the model with \emph{no} concentration dependence 
($\alpha=\beta=0$) for comparison. While this kind of model can reproduce the 
mean value, it fails to reproduce the variance.} 
Especially for the predicted variance, our model successfully
recovers the scatter in high-mass bins, where a model that depends
only on mass or the Poisson scatter cannot. For halos of the highest
and the lowest 25\% concentration in each mass bin, our model also
fits the simulation reasonably well.

In this work, we do not focus on refining these relations to obtain
the best mock subhalo abundance function. 
In fact, the essence of this work is to show that with our
simple model one can already reproduce most important features in the
subhalo abundance function.
\add{There are two main reasons for not pursuing the \emph{best-fit} model here.}

First of all, the parameterization proposed above is not unique.
For example, one can substitute the ratio $\Vmax/\Vvir$ that appears
in $V_0$ with some generic function of concentration $f(c)$, or put in
a mass/velocity dependency in $n$. The parameters can also involve
other host properties, or even be stochastic (i.e., involving random
variables).
\add{
Also, while the parameters provide insight on the dependence on concentration, they do not bear clear physical meaning and the parameterization choice is somewhat arbitrary.}

\add{Second, although simulations do provide constraints on the model parameters,
these parameters are very degenerate and the Poisson scatter of individual
halos makes it very difficult to tightly constrain the \emph{best-fit}
parameters. 
Multiple sets of values could give equally good fits to simulations}, 
and the choice of the objective function (statistics to minimize) would also 
affect the best-fit values.
The reported value in \autoref{table:para} are obtained by fitting only the mean and scatter of subhalo abundance in the full \texttt{c128-2048} box 
in bins of host halo mass \add{(i.e.~to minimize the two leftmost panels in \autoref{fig:model}), yet these values also provide decent fits to the individual abundance function as shown in \autoref{fig:slope}.}

\add{As a result, here we do not give meaningful 
error bars on the parameter values,
but rather simply demonstrate the model's capability of reproducing 
the subhalo abundance functions.}
Until the statistics of
high-resolution halos improves significantly, we recommend optimizing
the fit every time for each specific use case.

\subsection{The Power-law Index}

So far we have been fixing the power-law index (log--log slope) to be a
constant that does not change with halo properties when calibrating
our model against the \texttt{c125-2048} box.  This assumption is
consistent with previous studies~\citep[e.g.,][]{2012MNRAS.425.2169G}.
However, due to the resolution limit, low-mass host halos in a
cosmological box do not constrain the index as well as the high-mass
halos \add{because the number of resolved subhalos in low-mass host halos 
is smaller and subject to larger relative Poisson scatter. As a result,}
the value of $n_0$ in \autoref{table:para} is mostly set by
those high-mass halos in the box.

\begin{figure*}
  \centering \includegraphics[width=\textwidth]{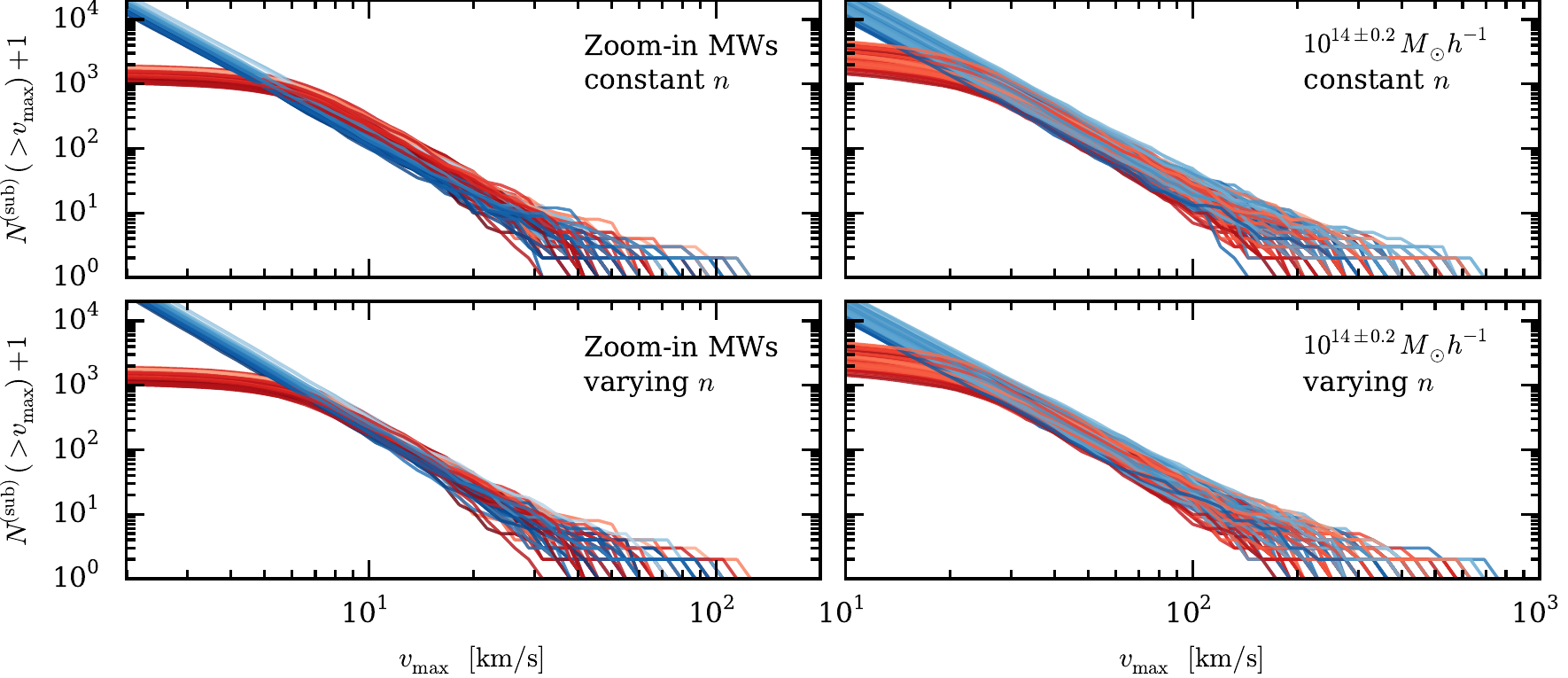}
  \caption{Subhalo abundance
    function in simulations (red) and predicted by the model (blue). The
    shade of colors represents the concentration ($\Vmax/\Vvir$)
    of the halo: the darker the more concentrated. The two columns show
    two different host halo masses. The upper row uses the model with
    constant index ($n=n_0$), while the lower row uses
    Equation~\eqref{eq:power_law}. The model with
    constant index cannot reproduce the subhalo abundance
    function for zoom-in Milky Way halos (upper left panel). \add{The gray band on the left shows the regime affected by resolution.}}
  \label{fig:slope}
\end{figure*}

To investigate whether the power-law index is indeed a constant,
we check if the model would work for both the zoom-in Milky Way halos and
the high-mass halos in the box.
In \autoref{fig:slope} we compare the subhalo abundance function
in simulations with that predicted by the model. We discover that a
constant index which can fit the subhalo abundance function very well
for cluster-size halos fails to fit the abundance function for zoom-in
Milky Way-size halos. The log--log slope of the abundance function is steeper
for Milky Way-size halos than for cluster-size halos.

We emphasize again that this mass trend is difficult to detect
in a cosmological box due to limited dynamical range. As shown in the
upper right panel of \autoref{fig:slope}, at $\vmax=50$ \kms{}, both
the number of subhalos and the scatter are still consistent with the
prediction from a constant slope.

Recall that the power-law index also changes with redshift, as shown
in \autoref{table:para}: at higher redshift, the log--log slope of the
abundance function is shallower.
The relation between the power-law index, host halo mass, and redshift
is also discussed in \citet{2005ApJ...624..505Z}, \citet{2011ApJ...738...22W}.
An intriguing question is then
whether this redshift trend and the aforementioned mass trend in the
index have the same physical origin.

Specifically, we find that we can fit the subhalo $\vmax$ functions
of the zoom-in Milky Way halos and of the cosmological box simultaneously
(see the lower panels of \autoref{fig:slope})
if we replace the constant index by this relation,
\begin{equation}
  n = -3.05\,\nu(M,z)^{-0.1},
  \label{eq:power_law}
\end{equation}
where
\[\nu(M, z) = \frac{\delta_c}{\sigma(M)D(z)}, \]
$\delta_c \approx 1.686$ is the critical overdensity,
$D(z)$ is the linear growth rate,
and $\sigma(M)$ is the squared root of the mass variance (at $z=0$)
with a top-hat filter of mass $M$.

\begin{figure}
  \centering \includegraphics[width=\columnwidth]{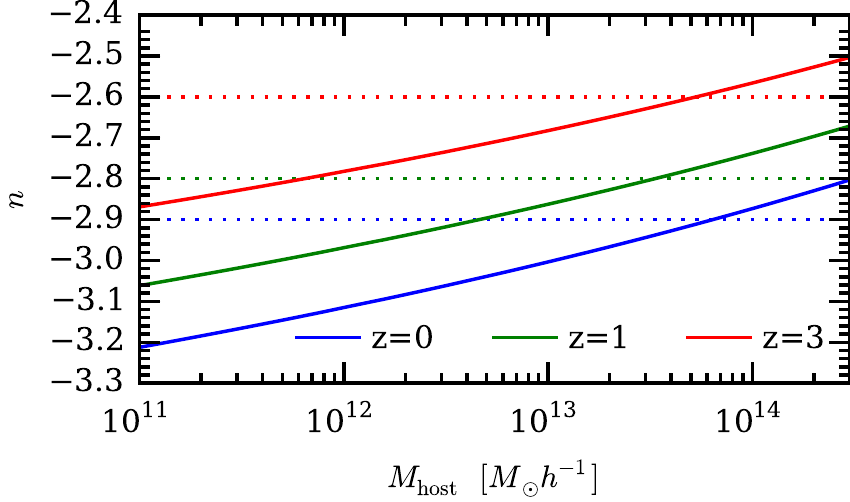}
  \caption{The solid lines show Equation~\eqref{eq:power_law}: the log--log
  slope as a function of mass at three different redshifts: $z=0$
  (blue), 1 (green), and 3 (red).  The dotted horizontal lines show
  the values of $n_0$ in \autoref{table:para} of corresponding
  redshifts.}
  \label{fig:slope_redshift}
\end{figure}

\autoref{fig:slope_redshift} shows the relation of Equation~\eqref{eq:power_law}
and compares it with the constant values of $n_0$ in \autoref{table:para}.
Although this is \textit{not} a proof of the validity of
Equation~\eqref{eq:power_law}, it indeed demonstrates the possibility
that the mass and redshift trends in the power-law index have the same
physical origin.  To robustly verify this connection between $n$ and
$\nu(M, z)$ would require several sets of zoom-in simulations of halos
of different masses, preferably also with different cosmologies.  This
is beyond the scope of this work, but worth exploring as simulation
suites expand.

\section{Implications and Discussion}
  \label{sec:discussion}

So far we have been focusing on \emph{subhalo} abundance function and
its dependence on host halo concentration.  In this section, we
discuss its observational implications.  While we cannot observe dark
matter subhalos directly, we can certainly count the satellite
galaxies that sit in those subhalos.  Hence, the \emph{subhalo}
occupation can be viewed as a proxy of the \emph{satellite}
occupation, subject to the effect of baryons on the
subhalo abundance function~\citep[e.g.,][]{2012MNRAS.423.2279C,2014MNRAS.444.1518V}.
Here we ignore baryonic effects and directly translate the subhalo occupation above a
certain velocity cut to the satellite abundance at a luminosity
threshold by specifying a galaxy--subhalo connection.

The simplest relation between subhalos and satellite galaxies is a one-to-one
relation,
\begin{equation}
\label{eq:one-to-one}
\nsub(>v) = \nsat(>L(v)),
\end{equation}
where $L(v)$ specifies the correspondence between velocity cut and
luminosity threshold by matching their abundance functions.  This is
commonly known as \emph{abundance
matching}~\cite[e.g.,][]{2004ApJ...609...35K,2004MNRAS.353..189V},
which has been shown to work fairly well for predicting measurements
such as the correlation
functions~\cite[e.g.,][]{2006ApJ...647..201C,2013ApJ...771...30R}.
With this abundance matching scheme, the model we introduced in
\autoref{sec:model} directly becomes $P(\nsat \vert M, c)$, and it
implies that satellite occupation depends on both host halo mass and
concentration.

A different, but also widely used approach is to use Halo Occupation
Distribution (HOD).  Instead of specifying the galaxy--subhalo
connection, standard HOD directly models the probability distribution
of satellite occupation at a luminosity threshold as a function of
host halo
mass~\citep[e.g.,][]{2000MNRAS.318.1144P,2000MNRAS.318..203S,2001ApJ...546...20S,2002ApJ...575..587B,2002PhR...372....1C}.
That is, it specifies $P(\nsat > L \vert M)$, and this distribution of
satellite occupation does not depend on host halo concentration.
Nevertheless, one can also generalize the HOD to include the
concentration dependence and to specify $P(\nsat \vert M, c)$.  Yet
most studies constraining HOD assume the sole dependence on mass.

Abundance matching and HOD also differ from each other in how the
positions of the satellite galaxies are assigned.  However, in the
context of satellite occupation, the only relevant difference is
whether or not the satellite occupation depends on host halo
concentration (at a given host halo mass).  It is clear that
\emph{subhalo} occupation does depend on host halo concentration, but
the stochastic process of galaxy formation could diminish this
dependence.  Nevertheless, it is also possible that
Equation~\eqref{eq:one-to-one} is only perturbed, and the concentration
dependence of \emph{subhalo} abundance still survives and results in
the concentration dependence of \emph{satellite} abundance.

In this section, we assume the simple relation of
Equation~\eqref{eq:one-to-one}, and investigate the implications of the
correlation between concentration and satellite occupation.  We
compare the different inferences between these two models (with and
without concentration dependence) when using satellite occupation as a
proxy of halo mass.  Then we look at the the possible signal of halo
assembly bias with satellite occupation.

\subsection{Satellite Occupation as a Proxy of Halo Mass}

Satellite occupation, especially in the cluster-mass regime, has been
used to probe the host halo
mass~\citep{2014MNRAS.441.1513O,2015MNRAS.449.1897O,2015ApJ...801...94O,2015MNRAS.450..592R}.
Conventionally, this is done within the standard HOD framework, which
ignores the dependence of satellite occupation on host halo
concentration.  Here we would like to investigate the effects of
ignoring this dependence.  We consider the two subhalo models, as
presented in \autoref{fig:model}: one only depends on halo mass like
the standard HOD, and the other incorporates the dependence on
concentration as introduced in \autoref{sec:model}.
We then take the host halos from simulations and populate them with subhalos 
according to these two models. 
This procedure is repeated multiple times to obtain enough statistics
and to smooth the Poisson noise.

\begin{figure}
  \centering \includegraphics[width=\columnwidth]{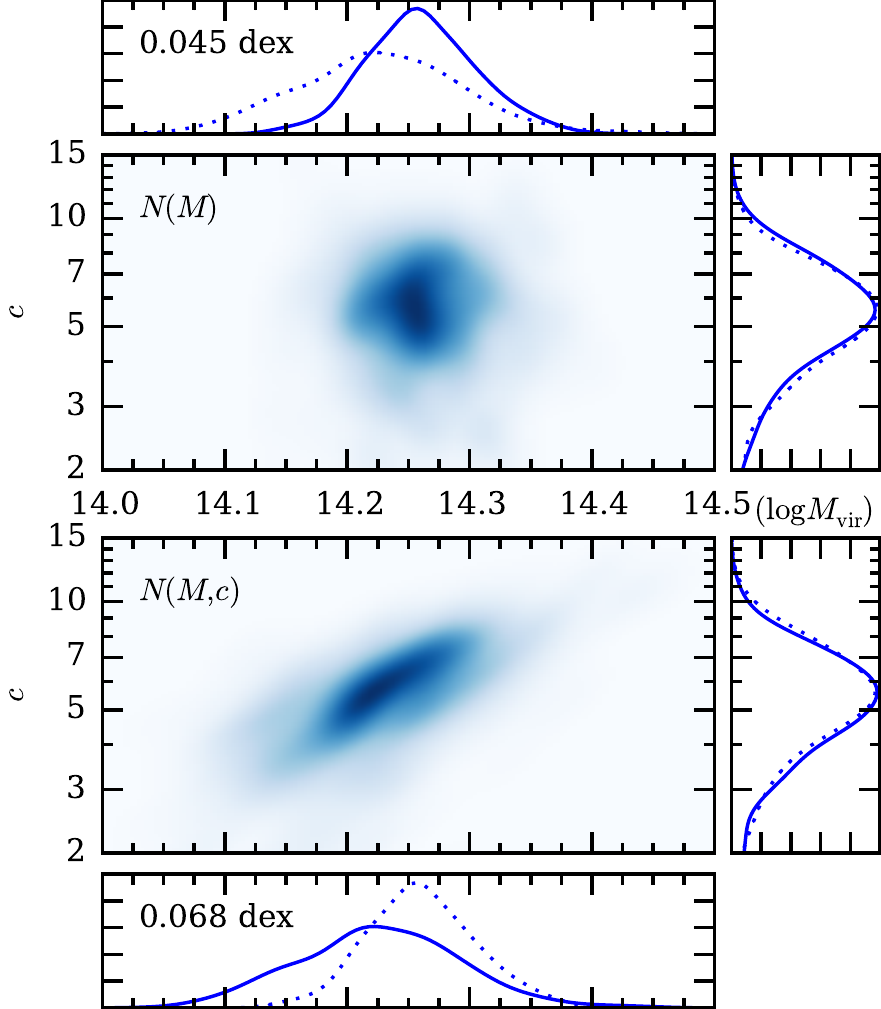}
  \caption{The joint and marginal distributions of logarithmic
  concentration ($y$-axis) and logarithmic mass ($x$-axis) of all the
  host halos which have exactly 100 subhalos whose $\vmax > 75$ \kms{}.
  The upper and lower parts demonstrate the inference from the two
  models: (1) with only mass dependence (upper) and (2) with both mass
  and concentration dependence (lower).  Dotted lines in the side
  panels show the same marginal distribution for the other model just
  for convenient comparison by eyes.  Both models are the same as used
  in \autoref{fig:model}. The number in the
  marginal distribution of logarithmic mass shows $\sigma$ value.}
  \label{fig:cm_fixedN}
\end{figure}

\autoref{fig:cm_fixedN} shows the joint distribution of the host halo
mass and concentration at a fixed satellite occupation, $\nsat(\vmax >
75~\text{km/s}) = 100$, in the context of cluster-size halos.  We see
significant differences between the inferences from the two subhalo
models, with or without the dependence on concentration.  Although the
mean value of inferred mass does not differ more than 1$~\sigma$, the
inferred distribution of mass is much wider in the case with the
dependence on concentration, and also includes many more
high-concentration high-mass or low-concentration low-mass halos.

The difference seen in \autoref{fig:cm_fixedN} would be especially
prominent when the number of subhalos in consideration is large
compared to the Poisson noise, i.e., $\nsat \gg \sqrt{\nsat}$.  Thus
when estimating the mass of galaxy clusters with richness or satellite
occupation, one should consider including halo concentration in the
model, especially in cases when not only the mean estimator but also
the resulting inference is relevant.

\begin{figure}
  \centering \includegraphics[width=\columnwidth]{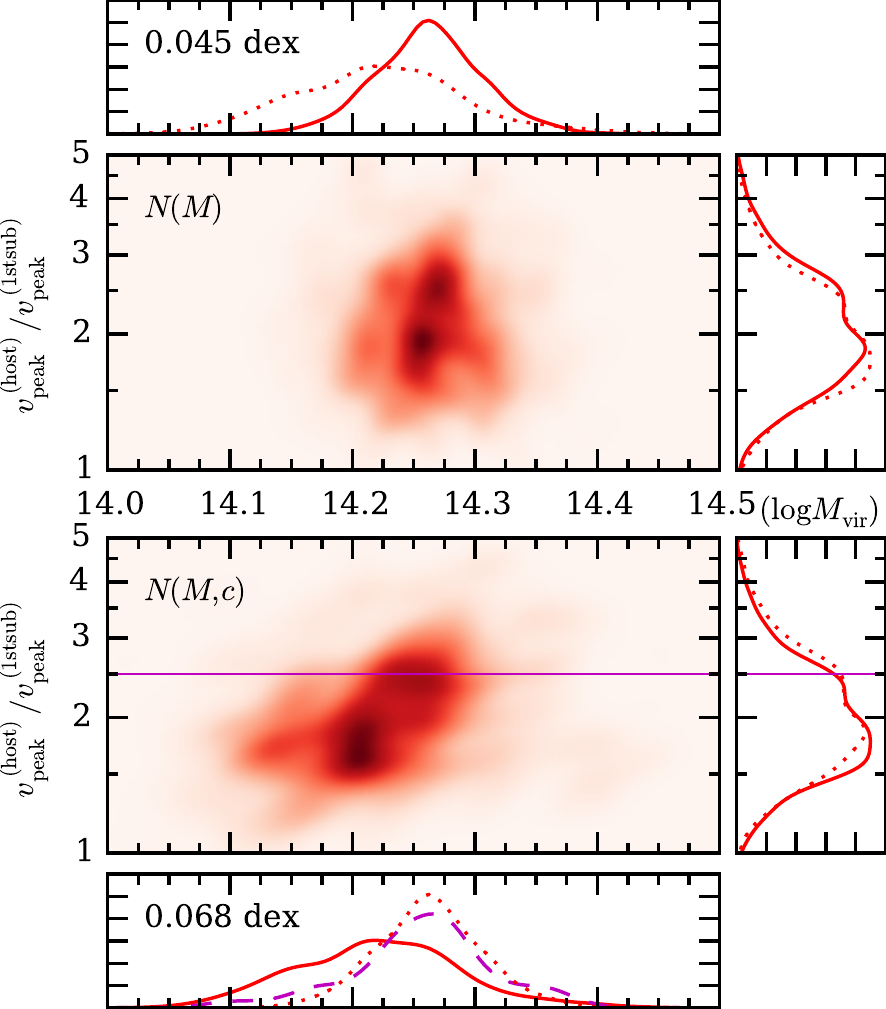}
  \caption{Same as \autoref{fig:cm_fixedN}, but showing the
  distributions of $\log \vpeak^\text{host}/\vpeak^\text{1st sub}$
  ($y$-axis) and logarithmic mass ($x$-axis).  The magenta dashed line
  in the lowest panel shows the mass distribution when selecting only
  halos whose ``gap'' is larger than 2.5.}
  \label{fig:gap-m_fixedN}
\end{figure}

To refine the mass estimator for halos of a fixed occupation,
we then need some independent observable to probe halo concentration.
We discuss three possible choices here.
\begin{enumerate}
\item \emph{The radial distribution of satellites.} If satellites trace the
density profile of the host halo, then by the radial distribution of satellites
could provide independent information on host halo concentration.
By comparing the number of satellites in different projected radial bins, 
one may be able to select those more concentrated halos in a fixed-richness sample.
\item \emph{The luminosity of the central galaxy.} 
For example, the abundance matching scheme of Equation~\eqref{eq:one-to-one} 
matches luminosity with $\vmax$ or $\vpeak$ instead of $\mvir$, and results in the
dependence of luminosity on concentration. 
Hence a further selection on the luminosity of central galaxy may provide 
a tighter mass distribution~\citep[see also][]{2008MNRAS.390.1157R}.
R.~M.~Reddick et al.~(2015, in preparation) also finds a negative correlation between 
the central luminosity and richness at a fixed halo mass, which agrees with trends
proposed here.
\item \emph{The magnitude gap}.
In addition to the concentration dependence of luminosity, 
the magnitude gap between the central galaxy and the brightest satellite galaxy
can further depend on the host halo concentration.
For instance, as suggested by our model, the parameter $\vcut$ itself has a
concentration dependence, regardless how luminosity is matched to halo properties.
It has also been shown in
simulations that the gap is correlated with the formation history of
the host halo, and hence with
concentration~\citep{2005ApJ...630L.109D,2005ApJ...624..505Z,2010MNRAS.405.1873D,2013ApJ...777..154D,2013ApJ...767...23W}.
\end{enumerate}

It has been suggested that selecting on magnitude gap can refine the
mass distribution of a fixed-richness
sample~\citep{2012ApJ...761..127M,2013MNRAS.430.1238H,2015ApJ...804...55L}.  Here we
revisit this method by considering the correlation between occupation
(richness) and halo concentration.  \autoref{fig:gap-m_fixedN} shows
the distributions of magnitude gap and halo mass, for a sample
of a fixed occupation (100 subhalos whose $\vmax > 75$ \kms{}, same as in
\autoref{fig:cm_fixedN}), for the two subhalo models.  
Here the magnitude gap is approximated by $\log
\vpeak^\text{host}/\vpeak^\text{1st sub}$, and can be 
translated into the actual magnitude map by abundance matching.
As we already learned, the distribution of halo mass is much wider (lower panel)
than that from the assumption that satellite occupation depends on
host halo mass only (upper panel).  Nevertheless, if we apply a
further selection on the magnitude gap, selecting only halos with
larger gaps, we can obtain a sample of halos whose mass distribution
is much closer to that in the upper panel.

This may provide a viable method to obtain a sample of halos in a narrower halo mass bin,
especially in the high-mass regime. 
It has been shown that selecting on magnitude
gap can indeed narrow the velocity dispersion distribution of the sample~\citep{2013MNRAS.430.1238H}.
As for halo mass, it remains to be seen how strong these effects are in specific observed samples, but we expect that the relative impact of the central galaxy luminosity and the magnitude gap could be tested in the near future using either lensing or X-ray measurements of large samples of optically selected clusters with fixed galaxy number.

\subsection{Satellites of Milky Way}

In the context of Milky Way-mass halos, 
the number of subhalos in consideration is much smaller,
and the Poisson noise of individual halos would dominate and diminish the
difference between these two subhalo models.  Nevertheless, in
\autoref{fig:cm_fixedN} we observe a positive correlation
between the host halo mass and concentration for this sample of a
fixed satellite occupation.  This positive correlation differs from
the commonly known concentration--mass
relation~\citep[e.g.,][]{1997ApJ...490..493N}, and can also been seen
when the number of subhalos in consideration is small.

\begin{figure}
  \centering \includegraphics[width=\columnwidth]{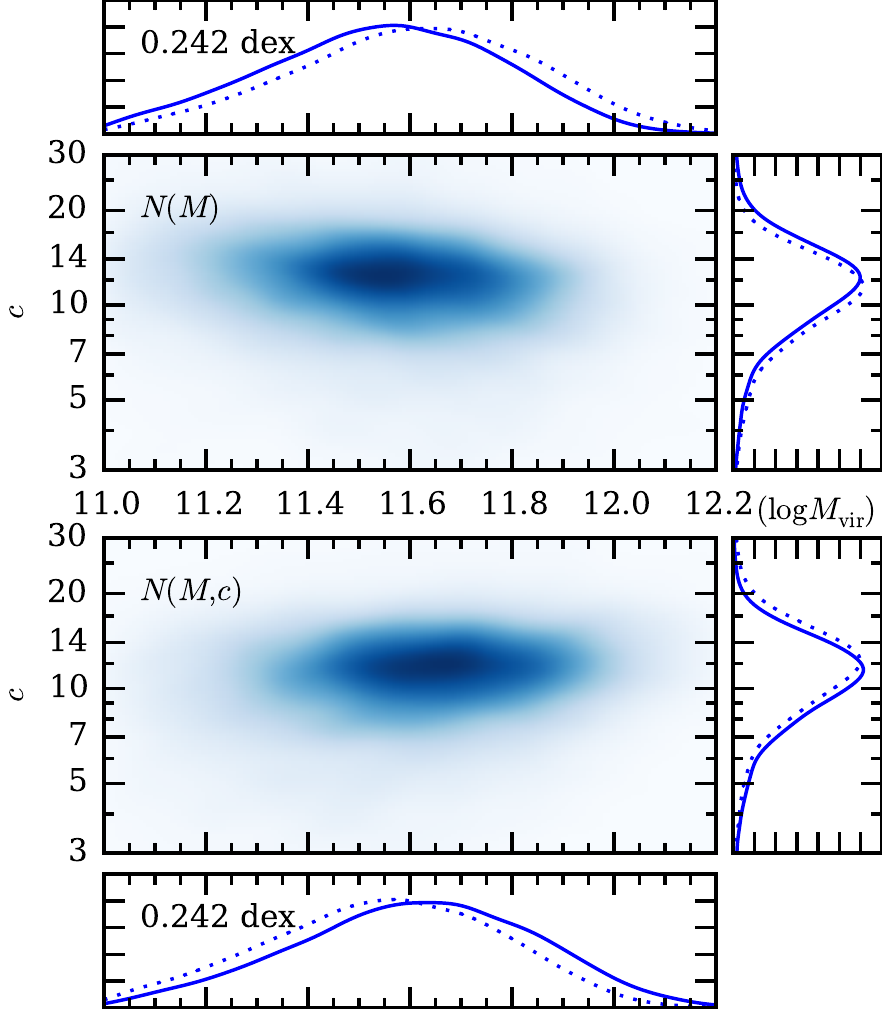}
  \caption{Same as \autoref{fig:cm_fixedN}, but showing of all the host halos
    which have exactly four subhalos whose $\vmax > 30$ \kms{}.}
  \label{fig:cm_fixedN_MW}
\end{figure}

\autoref{fig:cm_fixedN_MW} shows the joint distribution of the host halo mass and concentration at
another fixed satellite occupation, $\nsat(\vmax > 30~\text{ km/s}) =
4$.  In this case, the marginal distributions of mass or of
concentration barely differ between the two subhalo models.
Nevertheless, the predicted correlation between mass and concentration
is fairly different in the two cases.  Without the dependence on
concentration, a sample of a fixed satellite occupation basically
corresponds to a sample of halos in a mass bin, and the correlation
between halo concentration and mass inherits the usual, negative,
concentration--mass relation of host halos.  On the other hand, with
the dependence on concentration, the inferred correlation between
concentration and mass becomes positive.

This discrepancy again highlights the need to consider this dependence of
satellite occupation on concentration when inferring the mass or other properties of the
Milky Way halo from satellites~\citep[e.g.,][]{2011ApJ...743...40B,2013ApJ...773..172R,2013ApJ...767...92R,2014MNRAS.445.2049C}.
If the inference is not derived completely from simulations but 
with the help of a subhalo model which
does not account for dependence on concentration, such as the conventional
HOD, then one might need to consider the effect discussed above when
interpreting the results, particularly the degenerate correlation between 
concentration and mass.
We also note that recent constraint on the mass and concentration of the Milky Way from dynamical tracers
have a negatively correlated degeneracy~\citep{2015arXiv150203477W},
while occupation-based constraints will have the opposite degeneracy 
if the concentration dependence is properly accounted for, as demonstrated here.

This dependence on concentration also suggests that one should take 
the concentration of the Milky Way halo into account when investigating the tension between
the population of subhalos in $N$-body simulations and that of the observed Milky Way satellite galaxies~\citep[e.g.,][]{1993MNRAS.264..201K,1999ApJ...522...82K,1999ApJ...524L..19M,2010arXiv1009.4505B,2011MNRAS.415L..40B,2012JCAP...12..007P}.
While a Milky Way-like halo is conventionally defined by selecting on halo mass only, 
it is clear that the concentration of the Milky Way halo could potentially change 
the statistical significance of the aforementioned tension.
In a follow-up paper,
we further investigate these implications of this dependence on concentration
for the Milky Way and its population of satellites~(Y.-Y.~Mao et al.~2015, in preparation).

\subsection{Observing Halo Assembly Bias}

Given that satellite occupation is a direct observable that is
correlated with halo concentration, it may provide a way to
observationally detect the halo assembly bias.  Halo assembly bias has
been shown to exist in simulations; particularly it is found that host
halos of different formation histories or concentrations cluster
differently,
\begin{equation}
  b_\text{h}(M, c) \neq b_\text{h}(M),
  \label{eq:HAB}
\end{equation}
where $b_\text{h}$ is the halo bias
function~\citep{2005MNRAS.363L..66G,2006ApJ...652...71W,2007MNRAS.377L...5G}.
The question we want to address here is whether we can measure
\begin{equation}
  b_\text{h}(M, \nsat) \neq b_\text{h}(M),
  \label{eq:GAM_Nsub}
\end{equation}
and if so, whether it implies the existence of halo assembly bias
as in Equation~\eqref{eq:HAB}.

Instead of calculating the bias function directly,
we use the mark correlation function (MCF) to probe the bias.
The MCF is defined as
\begin{equation}
  \text{MCF}(m, r) = \sum_{(i,j) \in S_r}  \frac{m_i m_j}{\bar{m}^2},
\end{equation}
where $S_r = \{(i,j): \left\vert\mathbf{x}_i-\mathbf{x}_j\right\vert
\in [r,r+dr] \}$, and $\bar{m}$ is the mean of $m_i$ over $i$.  The
MCF of a specific mark $m$ shows whether the averaged value of
this mark for halos in pairs is higher or lower than the averaged
value of the whole sample.
\add{For each radial bin $S_r$, 
we find all pairs of halos whose separation falls
in that bin and measure the mark of those halos.}
To accommodate the possible large range of
the mark values, we use the ranks of the mark instead of the actual
value for $m$, normalized by the total number of different values.  If
Equation~\eqref{eq:GAM_Nsub} holds, we expect either a positive or a
negative excess in the MCF of $\nsat$.

In \citet{2006ApJ...652...71W}, the authors found a positive excess in
the MCF of $\nsat$ in the regime above $M_*$, but were not able to
find a similar signal below $M_*$.  To interpret these results, recall
that for halos below the typical collapse mass $M_*$,
high-concentrated halos are more clustered; for halos above $M_*$,
high-concentrated halos are less clustered.  In the regime above
$M_*$, halos in pairs are on averaged more massive but \emph{less}
concentrated, and both characters give a higher $\nsat$.  As a result,
the excess in the MCF of $\nsat$ comes from a mixed effect of both
mass and concentration, and hence it is easy to detect this excess but
would be difficult to distinguish whether this signal is really coming
from halo assembly bias.

On the other hand, in the regime below $M_*$, 
the dependence of the clustering strength on halo concentration switches sign,
but the dependence of $\nsat$ on concentration remains the same:
host halos that form earlier still have fewer subhalos at a fixed mass.
As a result, in the regime below $M_*$, halos in pairs are on
averaged more massive and \emph{more} concentrated, and these two
characters have opposite effects on $\nsat$.  If a negative excess in
the MCF of $\nsat$ is detected, this signal must come from the
contribution of concentration, or halo assembly bias.  However, in
\citet{2006ApJ...652...71W}, there were not enough subhalos resolved
in the simulation for the correlation between subhalo abundance and
halo concentration to manifest itself, and hence this signal was not
detected.

\begin{figure}
  \centering \includegraphics[width=\columnwidth]{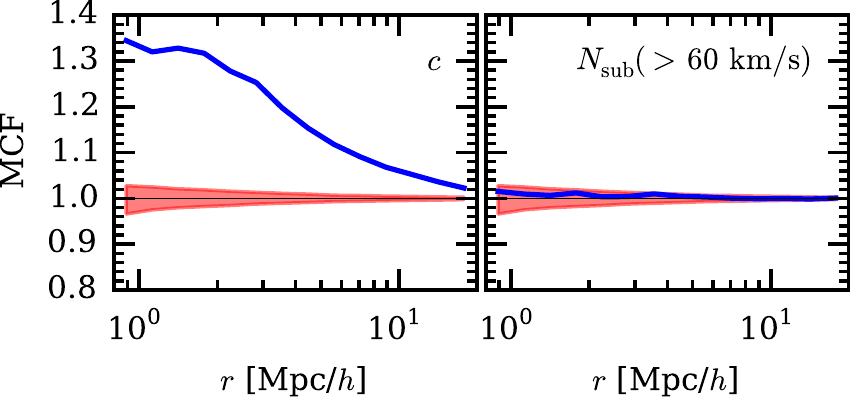}
  \caption{The MCFs of concentration
    (left) and of satellite occupation ($\vmax > 60$ \kms{}) (right),
    for host halos whose mass is within $10^{11}$ and $10^{11.4}~M_\odot h^{-1}$.
    The red shaded area shows the range of MCF consistent with no correlation
    within 2-$\sigma$.}
  \label{fig:mcf}
\end{figure}

We \add{first} calculate the MCFs of halo concentration and of satellite
occupation by selecting \add{all resolved} subhalos whose $\vmax > 60$ \kms{} \add{in our cosmological box}, for host
halos in a mass range, $10^{11}$--$10^{11.4}~M_\odot h^{-1}$, and plot
the results in \autoref{fig:mcf}.  The result we found here is
consistent with previous studies: significant bias in concentration,
but not in satellite occupation.  This result, however, does not
directly answer whether or not the satellite occupation can probe
assembly bias, because the variance in $\nsat$ can be large.  As we
argued in \autoref{sec:model},
\begin{equation}
  (\nsat \vert M, c) \sim \Pois(\langle\nsat \vert M, c\rangle).
\end{equation}
For host halos in this mass range, the number of resolved subhalos is
typically less than 10, even for a high-resolution cosmological box
(e.g., with a particle mass of $10^{7} M_\odot h^{-1}$). Despite the
correlation between subhalo abundance and host halo concentration, the
scatter in subhalo abundance can wash out this correlation, especially
for host halos with few subhalos, and render the bias in subhalo
occupation unobservable.

To verify our conjuncture that Equation~\eqref{eq:GAM_Nsub} would hold for
low-mass halos if the typical value of $\nsat$ is large ($> 10$), one
would need a cosmological box large enough to measure clustering
statistics and with a particle mass of $\sim 10^{5} M_\odot h^{-1}$,
but this kind of simulation is still beyond the reach of current
computational capabilities.  Zoom-in simulations can easily provide a
much better resolution, but those do not provide large-scale
statistics.  With our model, we can predict the expected number of
subhalos (satellites) to a lower velocity cut (higher number density),
while preserving the dependence on host halo mass and concentration.
We then can quantify at what velocity cut (number density) we can
start to observe the bias in subhalo occupation in low-mass host
halos.

\begin{figure}
  \centering \includegraphics[width=\columnwidth]{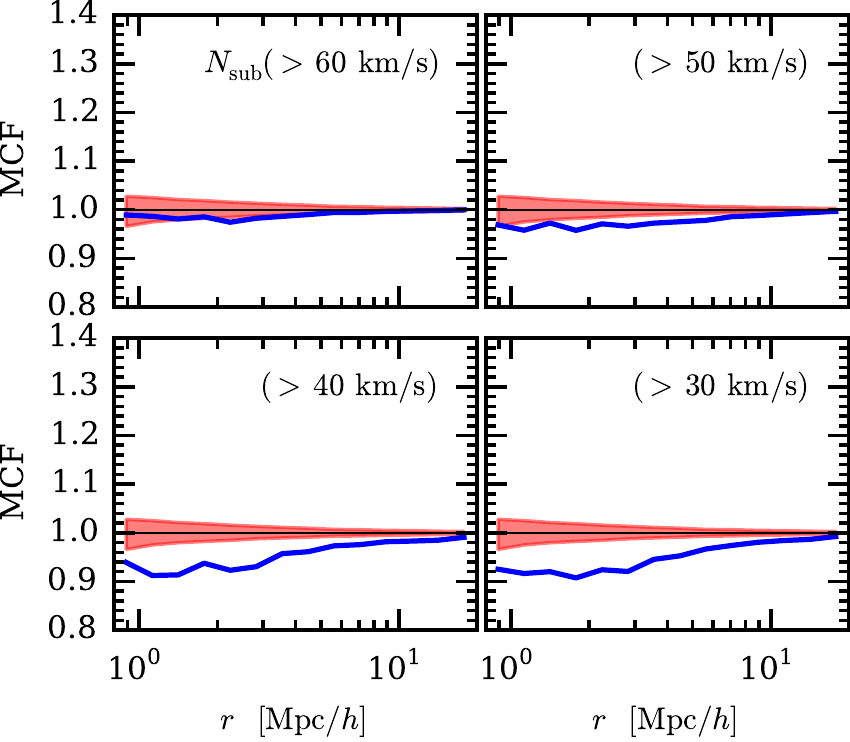}
  \caption{Same as \autoref{fig:mcf}, but shows the MCFs of
    \emph{model-predicted} satellite occupation down to 60, 50, 40, and 30 \kms{}.
    The corresponding number densities are 0.122, 0.216, 0.38,
    1.03~$(\text{Mpc}/h)^{-3}$.}
  \label{fig:mcf_model}
\end{figure}

\autoref{fig:mcf_model} shows the model-predicted MCF of subhalo
occupation for four different thresholds, in the same mass range of
the host halos, $10^{11}$--$10^{11.4}~M_\odot h^{-1}$. 
\add{The host halos are selected from the cosmological box, 
and for each host halo we re-populate its subhalos with our model.}
At $\vmax=60$ \kms{} the result can be directly compared with the right
panel of \autoref{fig:mcf}. Since our model by construction correlates
subhalo abundance and halo concentration ($\Vmax/\Vvir$), the lack of
signal in the MCF at $\vmax=60$ \kms{} results from the Poisson scatter.
Moving the threshold down to $\vmax=40$ \kms{} we start to see a clear
negative excess in the MCF.  As we discussed above, this negative
excess must originate from the fact that paired halos are on averaged
more concentrated, and hence have fewer subhalos.

\begin{figure}
  \centering \includegraphics[width=\columnwidth]{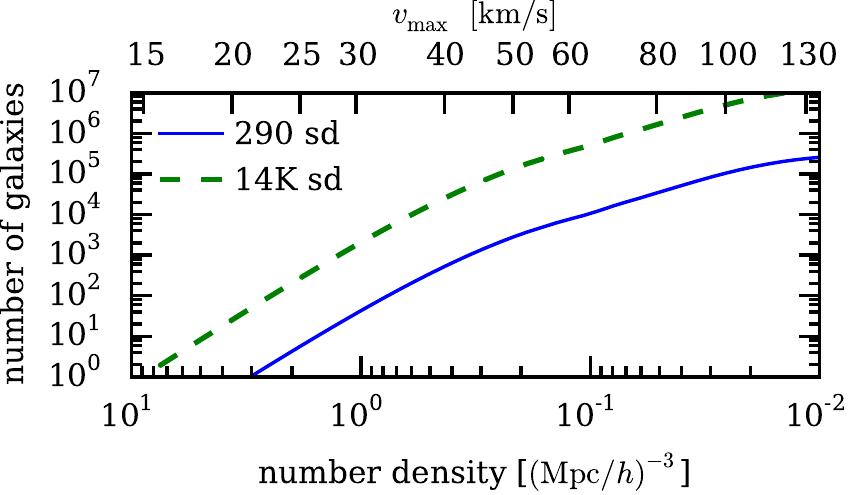}
  \caption{Expected number of galaxies in a volume-limited sample
    as a function of number density (and corresponding halo $\vmax$)
    for two example surveys with different sky coverage (given in square degrees).}
  \label{fig:survey}
\end{figure}

This negative excess in the MCF would manifest in the projected
correlation function by lowering the one-halo term if the
low-threshold data is available.  With upcoming deep spectroscopic surveys, such as
DESI~\citep{2013arXiv1308.0847L},
data with low thresholds will be accessible in the near future.
\autoref{fig:survey} demonstrates the number of galaxies in a
volume-limited sample from two exemplary surveys, assuming the
luminosity function reported in
\citet{2003ApJ...592..819B,2005ApJ...631..208B}.  Both surveys have a
detection limit of $m_r=19.5$, and their sky coverages are 290 and
14,000 square degrees, roughly representing the GAMA
Survey\footnote{\url{http://www.gama-survey.org/}}~\citep{2011MNRAS.413..971D}
and the DESI Bright Galaxy
Survey,\footnote{\url{http://desi.lbl.gov/cdr/}} respectively.  With
the latter survey, a volume-limited sample of a few hundred thousand
galaxies with $m_r<19.5$ down to the number density at
0.4~$(\text{Mpc}/h)^{-3}$ would be accessible, and this sample would
be sufficient for a precise measurement of the projected correlation
function.

We note that although we assume the simple relation of
Equation~\eqref{eq:one-to-one} in this discussion, this signal has the
advantage that it is less sensitive to the details of the galaxy--halo
relation because it only utilizes the number of satellites above a
certain luminosity threshold, but not other properties (e.g., color)
of the satellites.  Even if galaxy formation introduces additional
scatter in the satellite occupation, as long as this scatter is
smaller than the halo-to-halo scatter due to halo concentration, this
signal would survive in the projected correlation function.

\section{Summary}
  \label{sec:summary}

In this work, we model the subhalo abundance on the basis of
individual halos.  The framework of our model is based on the fact
that the scatter in $\nsub$ for an \emph{individual} halo is
consistent with Poisson scatter, and the additional halo-to-halo
scatter in $\nsub$ for halos \emph{in a mass bin} primarily affects
only the overall normalization of the subhalo function.  For a large
sample of halos, we find that the subhalo velocity functions of a sample of
halos in  a mass range are nearly parallel to one another. As a
result, we can model this halo-to-halo scatter by introducing
additional parameters to the model that specify the normalization as a
function of additional halo properties.

We hence present a model which predicts the subhalo abundance based on
two properties:  $\Vvir$ (equivalent to mass) and $\Vmax/\Vvir$
(roughly equivalent to concentration) of the host halos.  This model
successfully reproduces the mean and scatter in the subhalo abundance
in a given host halo mass bin. It can then be used to predict the
number of subhalos for thresholds that are lower than the resolution
limit of the simulation.  It also enables one to conveniently sample a
sequence of $\vmax$ values that represent the subhalos of a given host
halo.

This model further provides plain insight into the dependence of
subhalo abundance on halo concentration. We found that the halo
concentration affects the subhalo abundance function mainly through
the overall normalization  ($V_0$ in our parameterization), but also
through the ``cutoff'' scale ($\vcut$).  A constant power-law index
($n$) fits the cosmological simulations well; however, we also find
that an index that depends on halo mass would fit the zoom-in Milky Way halos
better. This dependence on mass may  have the same physical origin as
the dependence on redshift.

With this model, we then investigate the observable implications of the correlation between the subhalo abundance and halo concentration.
We find that when using subhalo or satellite occupation as a proxy of halo mass, one might need to consider using a
concentration-dependent model, such as the one presented here, to obtain a more accurate
inference.  We show that ignoring this dependence on concentration
could result in a biased mass inference and an incorrect joint
distribution of mass and concentration  of the sample.  Although these
biases are small, they may become important as other sources of
systematic errors decrease.

We further propose that satellite occupation can be used to probe halo
assembly bias if we can detect all satellites which reside in subhalos
down to $\sim 40$ \kms{}. Because in the low-mass regime,
high-concentration halos are more clustered but have fewer subhalos,
this signal can probe the halo assembly bias in concentration and is
\emph{not} degenerate with the contribution from halo mass.  This
method is also less sensitive to the detailed galaxy formation
processes because it only depends on the total count.

\acknowledgments{
Y.~Y.~M is supported by a Weiland Family Stanford Graduate Fellowship. M.~W.
received support from Stanford University grants for summer research.
This work was supported in part by the U.S. Department of Energy
contract to SLAC No.~DE-AC02-76SF00515. We thank Matthew Becker for
providing access to the cosmological simulations (\texttt{c125-2048}
and \texttt{c125-1024}) used in this work.  We thank Tom Abel, Matthew Becker,
Peter Behroozi, Michael Boylan-Kolchin, Andrew Hearin, Eduardo Rozo,
and Andrew Zentner for helpful discussions and comments.
Y.~Y.~M thanks CCAPP for its hospitality and for useful
discussion with participants at a Fall 2014 workshop on assembly bias.
The simulations used were run using computational resources at SLAC; we gratefully acknowledge the support of the SLAC computational team.
This research also used resources of the National Energy Research Scientific Computing Center, a DOE Office of Science User Facility supported by the Office of Science of the U.S. Department of Energy under Contract No.~DE-AC02-05CH11231.

}

\bibliographystyle{apj}
\bibliography{ref}

\end{document}